\documentclass[journal]{IEEEtran}
\ifCLASSINFOpdf
  \usepackage[pdftex]{graphicx}
  
  \DeclareGraphicsExtensions{.pdf,.jpeg,.png}
\fi

\usepackage{amsmath,bm}
\usepackage{fixmath}
\usepackage{amssymb}
\usepackage{textcomp}
\usepackage{algorithm,algcompatible,amsmath}

\usepackage{subfig}
\usepackage{gensymb}
\usepackage{textcomp}

\begin{document}

\title{A new approach for Weather Radars}

\author{Mohit~Kumar,~\IEEEmembership{Student Member,~IEEE}, V~Chandrasekar,~\IEEEmembership{Fellow,~IEEE}}

\markboth{}%
{Shell \MakeLowercase{\textit{et al.}}: Bare Demo of IEEEtran.cls for IEEE Journals}

\maketitle

\begin{abstract}
This paper elaborates the signal processing techniques for weather radars and their relative merits with respect to a similar phased array configuration. As will be shown in paper that this sub-aperture based configuration gives spatial resolution improvement compared to its phased array counterpart. This is the major benefit and a number of smaller benefits which are elaborated here for weather radar system. 
\end{abstract}


\IEEEpeerreviewmaketitle

\section{Introduction} \label{Intro}

\IEEEPARstart{T}{he} sub-array processing has been a concept which was lying around for years but now with ever increasing computing power, it is being realized. This type of radar can have multiple antenna elements in both transmit and receive, which looks like a phased array configuration. However, unlike phased arrays, it is able to transmit different waveforms that are orthogonal. The receive processing consists of a series of matched/mismatched based filters to separate out these echoes. \par
It can be superior to conventional phased arrays (electronic scanning) and the parabolic reflectors (mechanical scans) with better localization accuracy and angular resolution \cite{Bliss2003}. These can be attributed to the fact that due to the waveform diversity, the output signals of different sub-arrays appear as spatial samples corresponding to the convolution of the transmit and receive aperture phase centers, which produces additional virtual array elements (see \cite{Roberts2009} and \cite{Li2008}). The phase centers of the transmit and receive aperture in this sub-aperture radar system can be increased dramatically, which extends the array aperture \cite{Wang2008} and provides a potential for the system to have a higher spatial sample ability and space resolution than a conventional reflector antenna or a phased array system. \par
It has lot of potential for improving weather detection and enhancing the quality of polarimetric moments. This is because a sub-aperture array can potentially have a wider virtual array creation leading to higher spatial resolution. The higher spatial resolution is useful to identify some of the critical features in storms or tornadoes, where high resolution image can improve the quality of forecasting and help identifying such features. This concept was first developed for communications systems, for which orthogonal transmit-receive channels were used to circumvent channel fading \cite{Baker2013}. This type of configuration used widely separated antennas to exploit spatial diversity. This type of radar has better angle estimation accuracy compared to phased array radar because of an apparent decrease in antenna beamwidth when the linear array is operated as a sub-aperture radar. However, it is known that this type of a processing system suffers from interference due to parallel transmission of the probing signals (waveforms) which demands appropriate receiver to suppress the interference from other transmissions in the same radar. It can be observed that orthogonality is a key requirement to gain performance benefits claimed. The receiver and transmitter should ideally consist of perfect orthogonal waveforms with appropriate matched/mismatched filters. One such design using mismatched filters and polyphase orthogonal waveforms is presented in \cite{kumar2020}. It dwells on the design of a polyphase code pair and a mismatched filter pair that can achieve low level of peak auto-correlation and cross-correlation functions. A low level of peak cross-correlation function is essential to obtain orthogonality between the code/filter pairs. This can also be accomplished using the same waveform but modulated with different carrier frequency. \par
Another benefit in a weather radar system is the faster update rate possible because of synthesis of a twice broad transmit beam pattern on account of quadrant wise transmission (proposed here) and then using digital beamforming in receive to synthesize four simultaneous beams, two in azimuth and two in elevation direction. This can cut down the overall volume scan time by a fourth. The downside is the loss of directivity (and SNR) due to broad transmit pattern compared to a phased array of same antenna dimensions. The update rate is an advantage for weather events which are evolving fast and in general for weather forecasting and nowcasting systems.\par

\begin{figure}[!t]
	\centering
	\includegraphics[width=3in]{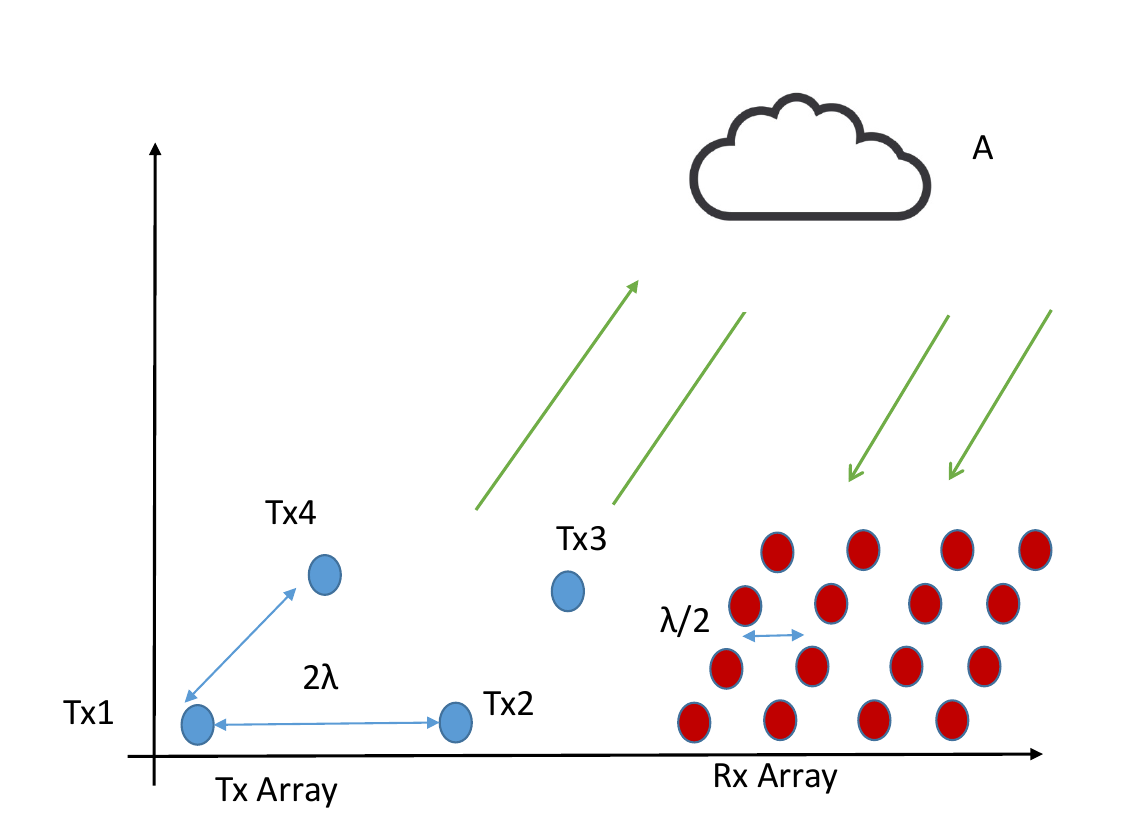}
	\caption{A conceptual representation of a weather sub-array radar configuration with M transmit and N receive antennas \cite{Ma2011}. }
	\label{fig_mimo}
\end{figure}

A simplified conceptual representation of a weather sub-aperture system is depicted in Fig. \ref{fig_mimo}. It is observing a weather event in the far-field and  is equipped with co-located antennas. Therefore, the directions of the target relative to the different transmit antennas and the receive antennas are the same, and the RCS of hydrometeors corresponding to different transmit–receive antenna pairs are also the same. The receive elements are in a 4x4 grid of $\lambda/2$ spacing and the corner elements are both transmit and receive. An important point to note here is that since the transmit elements are either energized one after the other (in case of time multiplexed operation) or they transmit individual orthogonal waveforms, each elemental pattern is omni-directional (because of no beamforming in transmit and beamforming happening only in receive). With this configuration, the weather radar would be able to see all directions instantaneously, an application very favorable for weather imaging. This is because all of the receive beams could be formed simultaneously by a 2-dimensional fast Fourier transform across the elements of the array. This weather imaging capability of a this radar can be very useful for rapidly evolving storms and tornadoes where high update rates of the event can be very beneficial to save thousands of lives by issuing warnings at the right time.

\section{Summary} 
The overall takeaway from this paper is that the sub-aperture based configuration is beneficial for weather radar as it improves the spatial resolution of the array without physical addition of more elements. But the price to pay is reduction in SNR and computational complexity of the system. \par

\bibliographystyle{IEEEtran}

\end{document}